# Deep learning of interface structures from the 4D STEM data: cation intermixing vs. roughening

M. P. Oxley,[1] J. Yin,[2] N. Borodinov,[1] S. Somnath,[2] M. Ziatdinov,[1,3] A. R. Lupini,[1]

S. Jesse,[1] R. K. Vasudevan*,[1] S. V. Kalinin,**[1]

[1]The Center for Nanophase Materials Sciences, Oak Ridge National Laboratory, Oak Ridge, TN 37831

[2]National Center for Computational Sciences, Oak Ridge National Laboratory, Oak Ridge, TN 37831

[3]Computational Sciences and Engineering Division, Oak Ridge National Laboratory, Oak Ridge, TN 37831

**Abstract**

Interface structures in complex oxides remain one of the active areas of condensed matter physics research, largely enabled by recent advances in scanning transmission electron microscopy (STEM). Yet the nature of the STEM contrast in which the structure is projected along the given direction precludes separation of possible structural models. Here, we utilize deep convolutional neural networks (DCNN) trained on simulated 4D scanning transmission electron microscopy (STEM) datasets to predict structural descriptors of interfaces. We focus on the widely studied interface between $LaAlO_3$ and $SrTiO_3$, using dynamical diffraction theory and leveraging high performance computing to simulate thousands of possible 4D STEM datasets to train the DCNN to learn properties of the underlying structures on which the simulations are based. We validate the DCNN on simulated data and show that it is possible (with >95% accuracy) to identify a physically rough from a chemically diffuse interface and achieve 85% accuracy in determination of buried step positions within the interface. The method shown here is general and can be applied for any inverse imaging problem where forward models are present.

*vasudevanrk@ornl.gov

**sergei2@ornl.gov

Oxide interfaces have emerged as one of the central topics in condensed matter physics research due to the multitude of unique behaviors they exhibit, ranging from interfacial conduction between dielectric materials,[18, 20] superconductivity, magnetic effects,[6, 9, 22] improper ferroelectric responses,[5] and large ionic conductivity.[10] This breadth of functional responses has stimulated outstanding theoretical efforts aimed at understanding the corresponding functionalities and their relationship with the interface structure. These discoveries have been enabled by the advances in oxide growth techniques including pulsed laser deposition, magnetron sputtering, and molecular beam epitaxy, often augmented by the reflection high energy electron diffraction (RHEED) as a control and monitoring tool that allows for the sub-unit cell precision.

A number of functional mechanisms for emergence of novel physical behaviors at the interfaces have been identified, ranging from classical semiconducting behaviors associated with mismatch between band offsets and resultant band bending, charge injection and changes in oxidation states, to symmetry mismatch across the interface and corresponding penetration of the distortions associated with zone center (e.g. ferroelastic and polarization) or zone boundary (e.g. octahedra tilts)[3, 12, 13] modes. For instance, a broad set of phenomena has been found to emerge at ferroelectric interfaces both due to the polarization screening and lattice symmetry breaking.[11] Combined with the field-induced switching, this resulted in significant interest to these material systems as potential multiferroics. However, the physical effects at the interface can compete with chemical interactions such as oxygen vacancy redistribution,[14, 15] compensating for the band offset effects[4] and polarization screening.[16] Many of these phenomena were explored and understood via recent advances in aberration-corrected scanning transmission electron microscopy (STEM), where the combination of the direct observations of atomic position enabled the reconstruction of order parameter fields and chemical strains, whereas electron energy loss spectroscopy (EELS) allowed insight into local charge states of cations and oxygen stoichiometry.

However, the fundamental limitation of STEM is the fact that the structural and chemical information is averaged along the beam path, limiting the separability of dissimilar interfacial mechanisms. One of the outstanding issues in this area is the potential cation mixing at the interfaces,[21] i.e. exchange between iso- or aliovalent cations in the adjacent layers during the growth. As can be readily understood, intermixing is expected to lead to effective doping in the adjacent layers, forming conductive channels, introducing new phases, etc. At the same time, in the projected image, intermixing will be indistinguishable from interface roughness, e.g. presence of substrate steps running across the interfaces, or island growth. Comparison of simulated and experimental electron energy loss spectroscopy (EELS) profiles have been used to determine if similar interfaces are abrupt.[7] However, if the experimental EELS profile across the interface is broader than expected, there is insufficient sensitivity in the EELS signal to differentiate between diffusion and interface roughness.

Here, we theoretically explore whether the cation intermixing and interface roughening can be distinguished using 4D STEM. In this technique, the 2D diffraction pattern from the sub-atomically focused electron beam is detected at each spatial location, giving rise to an information-rich 4D data set. However, the local diffraction pattern is determined both by the local material structure and the beam parameters including residual aberrations, spatial and temporal incoherence, obviating direct inversion. Here, we explore whether deep learning methods can be used to distinguish interface mechanisms from the 4D STEM data.

The main methodology used here is applicable to many inverse problems in imaging and is highlighted in Figure 1. Here, the range of physically realizable models of the material system corresponding to different classes of behavior or within a (small dimensional) parameter space are created, and their corresponding experimental fingerprints are modelled using a forward model. The corresponding model parameters or class are used as structural descriptors. A deep neural network is trained using the experimental fingerprints as an input and the structural descriptors as an output. Once trained on the corresponding training set, the network can be used on the unknown data to identify relevant structural descriptors. Notably, similar approach was implemented using back-propagation neural networks;[17, 19] however, poor generalizability and the *ad-hoc* approach to the formation of feature vectors precluded broad applications of this approach.

As a model system, we explored the most widely studied oxide interface in the recent past: i.e., the interface between the two band insulators $LaAlO_3$ (LAO) and $SrTiO_3$ (STO).[18, 24, 25] We began with structural models as indicated in Fig. 2(a-c). We model the system via a unit-cell-wide cross-section across the LAO-STO interface as shown in Fig. 2(a), with varying thicknesses in the [001] direction (typical of growth on (001) $SrTiO_3$ substrates). The STO substrate is terminated with a TiO2 layer, while the last LAO layer is terminated with an AlO layer. For maintaining periodic boundary conditions, a single SrO layer is added to cap the AlO. The section in Fig. 2(a) is what would be directly imaged by the electron beam. However, numerous structural possibilities exist in the direction of the beam, i.e. the [001] direction, including the existence of a buried step which causes roughening (as shown in Fig. 2(b)) and the B-site diffusion, i.e. of Ti into the Al sites. We therefore constructed dozens of possible structures where we varied the position of the buried interface, or alternatively, for a flat interface we varied the percentage of Ti substituting the Al in the first AlO layer at the interface (Fig. 2(c)). We then repeated these structures for varying thicknesses, from 100Å to 300Å. Once given the structures, we used the *μ*STEM simulation code[1, 2], and running on the OLCF's Titan supercomputer, to compute the convergent beam electron diffraction patterns from all the generated structures. The quantum excitation of phonons algorithm was used. Each CBED dataset was of size (341x341x13x123) In total the simulation size was on the order of ~350GB. The multislice routine was carried out using slices half the dimensions of the basic STO structure, with the total number of slices dependent on the simulated thickness. The structure was tiled 8 times in the y-direction to for the supercell and 8 configurations were calculated for each slice. This corresponds to a total of 64 configurations when the quick shifting option in *μ*STEM is used. Fifty passes were used at each probe position.

An example of the results of a simulation of a single structure is shown in Fig. 2(d), where we have computed the average intensity of each CBED pattern yield what would be roughly equivalent to a dark-field image. Since the dataset was large, and since the majority of the information is likely to reside at the interface, we restricted our study to the region shown by the bounded red lines in Fig. 2(d), which covers an area from the terminating TiO layer of the STO substrate to about one unit-cell into the LAO layer. To further reduce the size to enable deep learning, we then cropped the CBED patterns to only the central portions, as these are expected to contain the majority of the information. Note that wavelet compression could also be used in place of this step for forming a reduced representation, but we found it unnecessary. The CBED patterns along the dotted blue line in Fig. 2(d) are shown in Fig. 2(e). After cropping of the image sequence, one arrives at a single stack of images associated with that particular structure, and more such sets were generated via changing the location of the line profile (i.e., moving the line across the interface).

As a first step, we attempted to determine whether a deep convolutional neural network could take these sets of images and determine (classify) whether they originated from a rough interface (with a buried step), or from an interface that had some level of B-site diffusion. To do this, we used a deep convolutional neural network (DCNN) with 3D convolutional layers followed by an average pooling layer, and then two dense layers, followed by a final softmax layer for the classification for the task. The full architecture is shown in Fig. 3(a). It is worth mentioning that the reason for the lack of pooling (and especially max pooling) after each layer, which is typical for most DCNNs is that this results information loss on relative positions of features with respect to each other.[23] This information is important in diffraction and indeed we found that including such layers dramatically reduced the network accuracy. On the other hand, the lack of pooling also increases memory requirements. We also utilized dropout on each layer (15%) for reducing overfitting and trained the network on two nodes of the Summit supercomputer using the Horovod distributed framework, for our model built in Keras[8] using the Tensorflow backend. Although we had 600 simulations of each class, deep learning typically requires larger datasets to be effective. We therefore utilized data augmentation, via addition of varying levels of Poisson noise as well as small rotations, as can be expected in real experiments to artificially inflate the data volume. The data was validated on a set of simulated data that was not shown to the model during the training phase. We utilized standard SGD (lr = 0.01, momentum = 0.0) optimizer and trained for 300 epochs, resulting in a convergence of the model.

The results of the DCNN as a function of training can be seen in Fig. 3(b). After about 300 epochs the accuracy is approaches 98% on the validation set, suggesting that the network is almost perfectly capable of distinguishing between a rough interface from a chemically diffuse but sharp interface. An example of two predictions are shown in Fig. 3(c), where in the first case the DCNN predicts the image sequence shown contains a step (with >99% certainty, as gauged from the softmax layer output), and in the second case, as a chemically diffuse interface (with 100% from the softmax layer output), both of which are correct. Here we note that even to a skilled expert in CBED pattern analysis, distinguishing between the two cases is not straightforward. In fact, the network was trained on patterns derived from structures of different thickness, which greatly affects the resulting diffraction due to additional scattering as the probe propagates through the crystal.

Next, we explored whether we could utilize the same neural network architecture for a more challenging problem: for rough interfaces (those with buried steps), could the network be trained to determine where this buried interface resided? To simplify the problem somewhat, we resorted to attempting to determine the position of the buried step along the [001] axis as a percentage (to within 10%) of the thickness of the sample along the same direction. That is, if the step was 20% of the way up the sample, then the model should output 2, if 50% then 5, and so forth. We used the same network architecture as in Fig. 3(a) with the exception that the softmax layer had 11 outputs and used the same simulated data for training the network. Two changes had to be made in order for the network training to converge: (1) the training set had to be combined such that the CBED patterns had to be averaged across the unit cell, unlike the individual line profile in Fig. 2(d), and (2) we had to train separate models for each thickness, unlike in the previous case where a single model could be trained for all thicknesses. We are still exploring reasons for why this is so but are likely due to network capacity and the large variations in the CBED patterns as a function of thickness that make learning the buried step positions a very challenging task. The results of the training for two thicknesses are shown in Fig. 4(a,b). The DCNN appears to have an accuracy between 80-90% for the five thicknesses we explored (100 Å, 150 Å, 200 Å, 250 Å, 300 Å). Example predictions of the model for the four simulated CBED pattern series shown in Fig.4(c) shows that even when the prediction is incorrect (as in the second case), the predicted location is still close to the actual location of the buried step.

Here, we have shown that the use of hundreds of forward simulations combined with deep neural networks allows for extraction of structural descriptors. The method is general and can be applied for a range of imaging data wherever forward models are available. Extension to application to experimentally acquired CBED data is needed but will likely need both an order of magnitude more simulations to account for spatial and temporal decoherence effects of the electron beam that are present in real CBED patterns. We note that the code can run utilizing the GPU acceleration on SUMMIT's nodes, and in parallel fashion for different structural models, which makes such a strategy feasible computationally. Still, there may be distinct differences not captured within the simulation that are apparent in the experimentally observed patterns. Circumventing this may also be possible if we attempt to map the set of simulated patterns to the set of experimentally observed patterns via strategies such as cycle generative adversarial networks (cycle-GANs).[26]

In summary, we have shown using simulated data that 4D STEM can be a powerful tool in distinguishing between different structures buried within cross-section samples, via leveraging of high-performance computing and deep convolutional neural networks. We show that such neural networks have the ability to distinguish between chemically diffuse interfaces in LAO-STO, as opposed to physically rough interface formed via the inclusion of a buried step within the cross-sectional slice studied. In addition, we train a neural network with similar architecture to elucidate the location of the buried step, if it exists, and show that it is accurate on simulated data in ~85% of cases. The method here is applied to 4D STEM but can be useful in any imaging situation where multiple structural possibilities exist but the inversion of the acquired images is complex, provided that large scale simulations of the different structural possibilities is feasible.

**Acknowledgements**

The CBED simulations were supported by the U.S. Department of Energy (DOE), Office of Science (OS), Basic Energy Sciences, Materials Sciences and Engineering Division (MPO, SVK, ARL). The deep neural network research was part of the AI Initiative, sponsored by the Laboratory Directed Research and Development Program of Oak Ridge National Laboratory, managed by UT-Battelle, LLC, for the U.S. Department of Energy (DOE) (JY, NB, MZ, RKV, SJ). A portion of this research was conducted at the Center for Nanophase Materials Sciences, which is a US DOE Office of Science User Facility. This research used resources of the Compute and Data Environment for Science (CADES) at the Oak Ridge National Laboratory, which is supported by the Office of Science of the U.S. Department of Energy under Contract No. DE-AC05-00OR22725. The CBED simulations used resources of the Oak Ridge Leadership Computing Facility at the Oak Ridge National Laboratory, which is supported by the Office of Science of the U.S. Department of Energy under Contract No. DE-AC05-00OR22725.

**Figure Captions**

**Figure 1:** General idea of utilizing forward models in combination with deep learning to solve inverse problems in imaging. If a large number of forward simulations can be performed for different physically possible structures, then the output can be fed into a deep learning model and trained on this data to output the structural descriptors from which the forward simulation was performed. Then, given a new image, the deep learning model can predict the structural descriptor.

**Figure 2: Structure and CBED simulations. (a)** Idealized structure of the LAO-STO interface, one unit-cell wide. **(b)** Slice through the slab showing an example structure with a buried step. **(c)** Slice through the first Al layer with 20% of the Al atoms randomly replaced by Ti atoms. **(d)** Mean intensity of CBED diffraction patterns from one simulation (roughly equivalent to HAADF image). Each pixel in this image has an associated CBED pattern, some of which are shown in **(e)**, where the CBED patterns arising from the line profile in (d) are plotted. This image sequence is cropped to the center as shown in the figure, and the resulting cropped image stack comprises one training example for the deep neural network. About 1200 such training examples were formed, based on different buried step positions [see (b)] and varying the amount and randomizing the location of the B-site diffusion [see (c)].

**Figure 3:** Deep convolutional neural network architecture and results on rough interface vs. chemically diffuse interface. **(a)** Network architecture, which consists of several 3D convolutional layers and two dense layers before a final softmax classification layer. Dropout is used on each layer. The kernel size in the convolutional layers is indicated. **(b)** Training and validation accuracy during training of the network. **(c)** Example predictions on the validation set.

**Figure 4:** Step position determination network. Results during training of the network for (a) 150 Å and (b) 250 Å. (c) Example predictions of step positions and their actual positions ('Actual'), for slabs of thickness 250 Å.

Figure 1

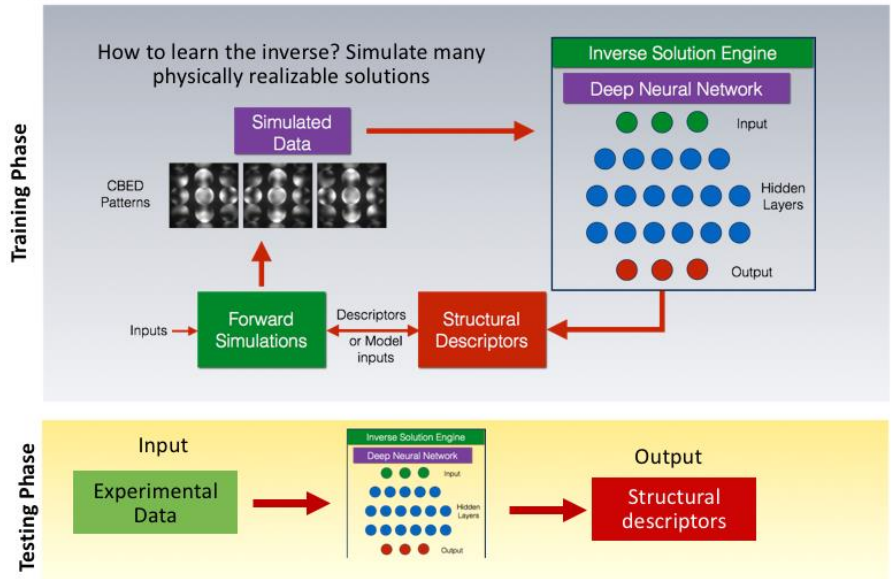

Figure 2

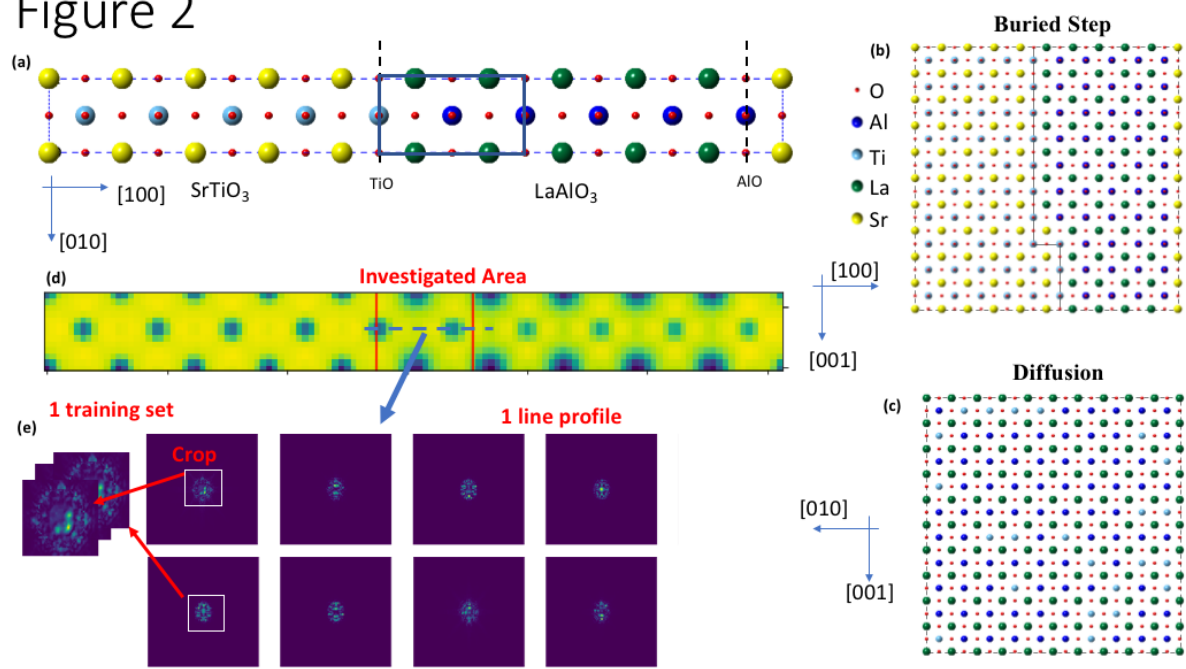

Figure 3

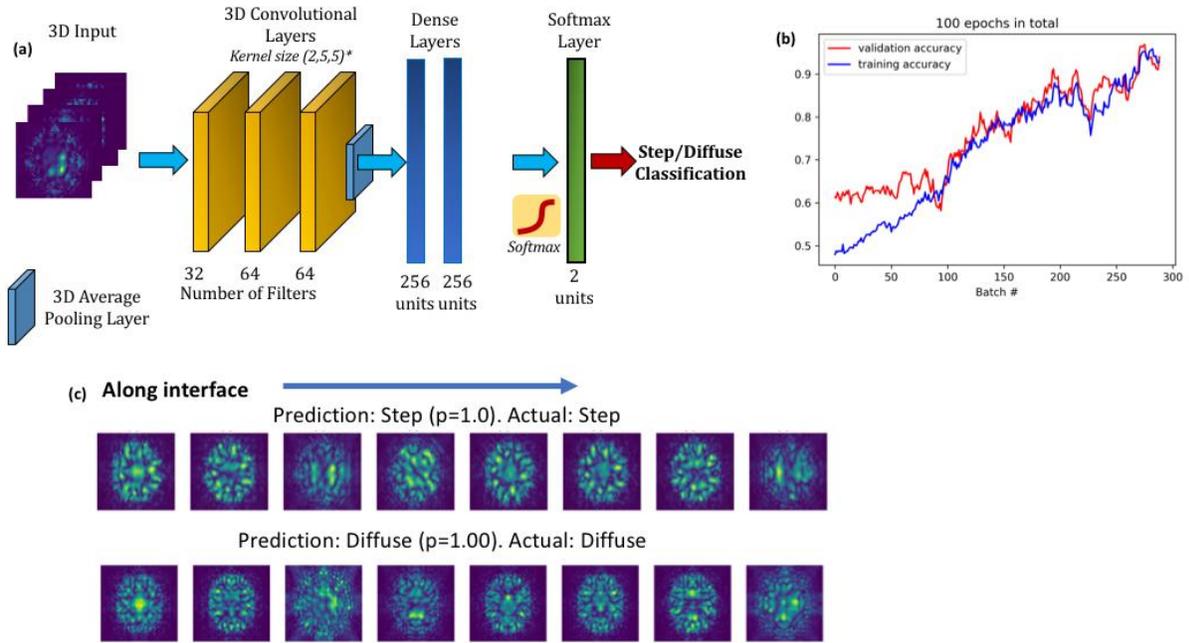

Figure 4

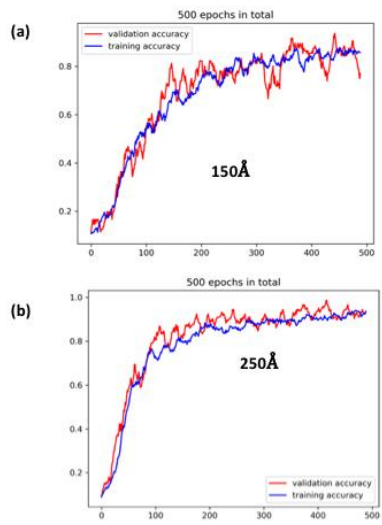
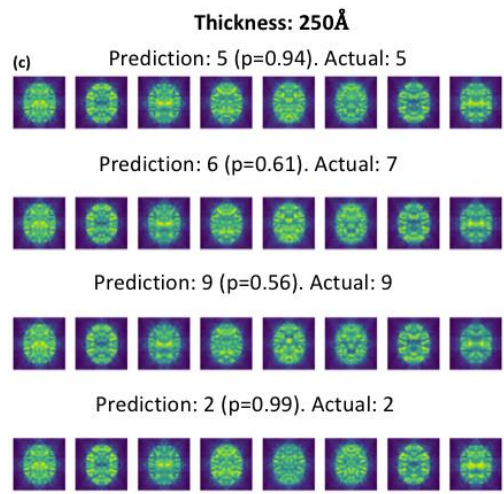